
\def\ung{{{\frak{g}}}}
\def\ungh{{{{\hat{\frak{g}}}}}}
\def\uqg{{{U_{q}(\ung)}}}
\def\uqgh{{{U_{q}(\ungh)}}}

\def\bp{{{\bold{P}}}}
\def\bq{{{\bold{Q}}}}
\def\calp{{{{\calp}}}}

\def\ot{{{\otimes}}}
\def\op{{{\oplus}}}

\def\unh{{{\frak{h}}}}

\def\calp{\Cal P}

\def\vbp{v_{\bp}}
\def\vbq{v_{\bq}}

\magnification 1200

\input amstex
\NoBlackBoxes

\documentstyle{amsppt}
\document
\centerline{\bf{{ Minimal Affinizations of Representations}}}
\vskip 12pt
\centerline{\bf of Quantum Groups:}
\vskip 12pt
\centerline{\bf the $U_q(\ung)$--module structure}
\vskip 36pt
\centerline{Vyjayanthi Chari{\footnote{Partially supported by the NSF,
DMS--9207701}},}
\vskip 12pt
\centerline{Department of Mathematics,}
\vskip 12pt
\centerline{University of California, Riverside, CA 92521, USA.}
\vskip 24pt

\vskip 36pt
{\eightpoint{{\centerline{Abstract}}
\noindent If $U_q(\frak g )$ is a
finite--dimensional complex simple Lie algebra,
an affinization of a finite--dimensional irreducible representation $V$ of
$\uqg$ is a finite--dimensional irreducible representation $\hat{V}$ of $\uqgh$
which contains $V$ with multiplicity one, and is such that all other
$\uqg$--types in $\hat{V}$ have highest weights strictly smaller than that of
$V$.
There is a natural partial ordering $\preceq$ on the set of affinizations of
$V$ defined in [2]. If $\ung$ is of rank 2, we prove in [2] that there is
unique minimal element with repsect to this order. In this paper, we give the
$U_q(\ung)$--module structure of the minimal affinization when $\ung$ is of
type $B_2$.}}
\vskip 36pt

\noindent{\bf Introduction}
\vskip12pt\noindent In [2], we defined the notion of an
affinization of a finite-dimensional irreducible representation $V$ of the
quantum group $\uqg$, where $\ung$ is a finite-dimensional complex simple Lie
algebra and $q\in\Bbb C^\times$ is transcendental. An affinization of $V$ is an
irreducible representation $\hat{V}$ of the quantum affine algebra $\uqgh$
which, regarded as a representation of $\uqg$, contains $V$ with multiplicity
one, and is such that all other irreducible components of $\hat{V}$ have
highest weights strictly smaller than that of $V$. We say that two
affinizations are equivalent if they are isomorphic as representations of
$\uqg$. We refer the reader to the introduction to [2] for a discussion of the
significance of the notion of an affinization.

An interesting problem is to describe the structure of $\hat{V}$ as a
representation of $\uqg$. This problem appears difficult for an arbitrary
affinization; however, in [2] we introduced a partial order on the set of
equivalence classes of affinizations of $V$ and proved that there is a unique
minimal affinization if $\ung$ is of rank 2.  If $\ung$ is of type $A$, it was
known that every $V$ has an affinization $\hat{V}$ which is irreducible under
$\uqg$; it was proved in [4] that $\hat{V}$ is the unique minimal affinization
up to equivalence.
However, if $\ung$ is not of type $A$, there is generally no affinization of a
given representation $V$ which is irreducible under $U_q(\ung)$ and the
description of the structure of the  minimal affinizations as representations
of $\uqg$ is not obvious. Some examples were worked out in [7]; in this paper,
we describe the $\uqg$-structure of the minimal affinization of an arbitrary
irreducible representation of $V$ when $\ung$ is of type $B_2$.
A consequence of our results is that the minimal affinizaton of $V$ is
irreducible under $\uqg$ if and only if the value of the highest weight on the
short simple root of $\ung$ is 0 or 1.

\vskip36pt\noindent{\bf 1 Quantum affine algebras and their representations}
\vskip 12pt\noindent
In this section, we collect the results about quantum affine algebras which we
shall need later.

Let $\ung$ be a finite--dimensional complex simple Lie algebra with Cartan
subalgebra $\unh$ and Cartan matrix $A=(a_{ij})_{i,j\in I}$. Fix coprime
positive integers  $(d_i)_{i\in I}$\/ such that $(d_ia_{ij})$\/ is symmetric.
Let $P=\Bbb Z^I$ and let
$P^+=\{\lambda\in P\mid \lambda(i)\ge 0\ \text{for all $i\in I$}$\}. Let $R$
(resp. $R^+$) be the set of roots (resp. positive roots) of $\ung$. Let
$\alpha_i$ ($i\in I$) be the simple roots and let $\theta$ be the highest root.
Define a non-degenerate symmetric bilinear form $(\, , \, )$ on $\unh^*$ by
$(\alpha_i,\alpha_j)=d_ia_{ij}$,
and set $d_0=\frac12(\theta,\theta)$. Let $Q = \op_{i\in I}\Bbb
Z.\alpha_i\subset\unh ^*$\/ be the root lattice, and set $Q^+ =\sum_{i\in
I}\Bbb N.\alpha_i$. Define a partial order $\ge$ on $P$ by $\lambda\ge \mu$ iff
$\lambda-\mu\in Q^+$. Let $\lambda_i$ ($(i\in I)$) be the fundamental weights
of $\ung$, so that $\lambda_i(j) =\delta_{ij}$.

In this paper, we shall be interested in the case when $\ung$ is of type $B_2$.
Then,
$$ \align
I=\{1,2\},\ \ \ \ \ d_0&=d_1 =2,\ \  d_2 =1,\ \ \theta=\alpha_1+2\alpha_2,\\
A & = \left(\matrix 2&-1\\-2&2\endmatrix\right).
\endalign$$

Let $q\in \Bbb C^{\times}$ be transcendental, and, for $r,n\in\Bbb N$, $n\ge
r$, define
$$\align [n]_q & =\frac{q^n -q^{-n}}{q -q^{-1}},\\
[n]_q! &=[n]_q[n-1]_q\ldots [2]_q[1]_q,\\
\left[{n\atop r}\right]_q &= \frac{[n]_q!}{[r]_q![n-r]_q!}.\endalign$$

\proclaim{Proposition 1.1} There is a Hopf algebra $\uqg$ over $\Bbb C$ which
is generated as an algebra by elements $x_i^{{}\pm{}}$, $k_i^{{}\pm 1}$ ($i\in
I$), with the following defining relations:
$$\align
k_ik_i^{-1} = k_i^{-1}k_i &=1,\;\;  k_ik_j =k_jk_i,\\
k_ix_j^{{}\pm{}}k_i^{-1} &= q_i^{{}\pm a_{ij}}x_j^{{}\pm},\\
[x_i^+ , x_j^-] &= \delta_{ij}\frac{k_i - k_i^{-1}}{q_i -q_i^{-1}},\\
\sum_{r=0}^{1-a_{ij}}\left[{{1-a_{ij}}\atop r}\right]_{q_i}
(x_i^{{}\pm{}})^rx_j^{{}\pm{}}&(x_i^{{}\pm{}})^{1-a_{ij}-r} =0, \ \ \ \ i\ne
j.\endalign$$

The comultiplication $\Delta$, counit $\epsilon$, and antipode $S$ of $\uqg$
are given by
$$\align\Delta(x_i^+)&= x_i^+\ot k_i +1\ot x_i^+,\\
\Delta(x_i^-)&= x_i^-\ot 1 +k_i^{-1}\ot x_i^-,\\
\Delta(k_i^{{}\pm 1}) &= k_i^{{}\pm 1}\ot k_i^{{}\pm 1},\\
\epsilon(x_i^{{}\pm{}}) =0,\;\ & \epsilon(k_i^{{}\pm 1}) =1,\\
S(x_i^+) = -x_i^+k_i^{-1},\; S(x_i^-) &=- k_ix_i^-, \; S(k_i^{{}\pm 1})
=k_i^{{}\mp 1},\endalign$$
for all $i\in{I}$.\qed
\endproclaim

Let $\hat{I} = I\amalg\{0\}$\/ and let ${\hat A} =(a_{ij})_{i,j\in {\hat I}}$
be the extended Cartan matrix of $\ung$, i.e. the generalized Cartan matrix of
the (untwisted) affine Lie algebra $\hat\ung$ associated to $\ung$. Let
$q_0=q^{d_0}$.

When $\ung$ is of type $B_2$,
$$\hat{A}= \left(\matrix 2&0&-1\\0&2&-1\\-1&-2&2\endmatrix\right).$$

\proclaim{Theorem 1.2} Let $\uqgh$ be the algebra with generators
$x_i^{{}\pm{}}$, $k_i^{{}\pm 1}$ ($i\in\hat{I}$) and defining relations those
in 1.1, but with the indices $i$, $j$ allowed to be arbitrary elements of $\hat
I$. Then, $\uqgh$ is a Hopf algebra with comultiplication, counit and antipode
given by the same formulas as in 1.1 (but with $i\in\hat{I}$).

Moreover, $\uqgh$ is isomorphic to the algebra ${\Cal A}_q$ with generators
$x_{i,r}^{{}\pm{}}$ ($i\in I$, $r\in\Bbb Z$), $k_i^{{}\pm 1}$ ($i\in I$),
$h_{i,r}$ ($i\in I$, $r\in \Bbb Z\backslash\{0\}$) and $c^{{}\pm{1/2}}$, and
the following defining relations:
$$\align
c^{{}\pm{1/2}}\ &\text{are central,}\\
k_ik_i^{-1} = k_i^{-1}k_i =1,\;\; &c^{1/2}c^{-1/2} =c^{-1/2}c^{1/2} =1,\\
k_ik_j =k_jk_i,\;\; &k_ih_{j,r} =h_{j,r}k_i,\\
k_ix_{j,r}k_i^{-1} &= q_i^{{}\pm a_{ij}}x_{j,r}^{{}\pm{}},\\
[h_{i,r} , x_{j,s}^{{}\pm{}}] &= \pm\frac1r[ra_{ij}]_{q_i}c^{{}\mp
{|r|/2}}x_{j,r+s}^{{}\pm{}},\\
x_{i,r+1}^{{}\pm{}}x_{j,s}^{{}\pm{}} -q_i^{{}\pm
a_{ij}}x_{j,s}^{{}\pm{}}x_{i,r+1}^{{}\pm{}} &=q_i^{{}\pm
a_{ij}}x_{i,r}^{{}\pm{}}x_{j,s+1}^{{}\pm{}}
-x_{j,s+1}^{{}\pm{}}x_{i,r}^{{}\pm{}},\tag1\\
[h_{i,r},h_{j,s}]&=\delta_{r,-s}\frac1{r}[ra_{ij}]_{q_i}\frac{c^r-c^{-r}}{q_j-q_j^{-1}},\\
[x_{i,r}^+ , x_{j,s}^-]=\delta_{ij} & \frac{ c^{(r-s)/2}\phi_{i,r+s}^+ -
c^{-(r-s)/2} \phi_{i,r+s}^-}{q_i - q_i^{-1}},\endalign$$
$$
\sum_{\pi\in\Sigma_m}\sum_{k=0}^m(-1)^k\left[{m\atop k}\right]_{q_i} x_{i,
r_{\pi(1)}}^{{}\pm{}}\ldots x_{i,r_{\pi(k)}}^{{}\pm{}}  x_{j,s}^{{}\pm{}}
 x_{i, r_{\pi(k+1)}}^{{}\pm{}}\ldots x_{i,r_{\pi(m)}}^{{}\pm{}} =0,
\tag2$$
if $i\ne j$, for all sequences of integers $r_1,\ldots, r_m$, where $m
=1-a_{ij}$, $\Sigma_m$ is the symmetric group on $m$ letters, and the
$\phi_{i,r}^{{}\pm{}}$ are determined by equating powers of $u$ in the formal
power series
$$\sum_{r=0}^{\infty}\phi_{i,\pm r}^{{}\pm{}}u^{{}\pm r} = k_i^{{}\pm 1}
exp\left(\pm(q_i-q_i^{-1})\sum_{s=1}^{\infty}h_{i,\pm s} u^{{}\pm s}\right).$$

If $\theta =\sum_{i\in I}m_i\alpha_i$, set $k_{\theta} = \prod_{i\in
I}k_i^{m_i}$. Suppose that the root vector $\overline{x}_{\theta}^+$ of $\ung$
corresponding to $\theta$ is expressed in terms of the simple root vectors
$\overline{x}_i^+$ ($i\in I$) of $\ung$ as
$$\overline{x}_{\theta}^+ = \lambda[\overline{x}_{i_1}^+, [\overline
x_{i_2}^+,\cdots ,[\overline x_{i_k}^+, \overline x_j^+]\cdots ]]$$
for some $\lambda\in\Bbb C^{\times}$. Define maps $w_i^{{}\pm{}}:\uqgh\to\uqgh$
by
$$w_i^{{}\pm{}}(a) = x_{i,0}^{{}\pm{}}a - k_i^{{}\pm 1}ak_i^{{}\mp
1}x_{i,0}^{{}\pm{}}.$$
Then, the isomorphism $f:\uqgh\to\Cal A_q$ is defined on generators by
$$\align
f(k_0) = k_{\theta}^{-1}, \ f(k_i) &= k_i, \ f(x_i^{{}\pm{}}) =
x_{i,0}^{{}\pm{}},  \ \ \ \ (i\in I),\\
f(x_0^+) &=\mu w_{i_1}^-\cdots w_{i_k}^-(x_{j,1}^-)k_{\theta}^{-1},\\
f(x_0^-) &=\lambda k_{\theta} w_{i_1}^+\cdots w_{i_k}^+(x_{j,-1}^+),\endalign
$$
where $\mu\in\Bbb C^{\times}$ is determined by the condition
$$[x_0^+, x_0^-] =\frac{k_0-k_0^{-1}}{q_0-q_0^{-1}}. \qed$$
\endproclaim

See [1], [5] and [9] for further details.

Note that there is a canonical homomorphism $\uqg\to\uqgh$ such that
$x_i^{{}\pm{}}\mapsto x_i^{{}\pm{}}$, $k_i^{{}\pm 1}\mapsto k_i^{{}\pm 1}$ for
all $i\in I$. Thus, any representation of $\uqgh$ may be regarded as a
representation of $\uqg$.

It is easy to see that, for any $a\in\Bbb C^\times$, there is a Hopf algebra
automorphism $\tau_a$ of $\uqgh$ given by
$$\align
\tau_a(x_{i,r}^{{}\pm{}})=a^rx_{i,r}^{{}\pm{}},\
&\tau_a(\phi_{i,r}^{{}\pm{}})=a^r\phi_{i,r}^{{}\pm{}},\\
\tau_a(c^{\frac12})=c^{\frac12},\ &\tau_a(k_i)=k_i,\endalign$$
for $i\in I$, $r\in \Bbb Z$ (see [5]).

Let $\hat U^{{}\pm{}}$ (resp. $\hat U^0$) be the subalgebra of  $\uqgh$
generated by the $x_{i,r}^{{}\pm{}}$ (resp. by the $\phi_{i,r}^{{}\pm{}}$) for
all $i\in I$, $r\in\Bbb Z$. Similarly, let $U^{{}\pm{}}$ (resp. $U^0$) be the
subalgebra of $\uqg$ generated by the $x_i^{{}\pm{}}$ (resp. by the $k_i^{{}\pm
1}$) for all $i\in I$.
\proclaim{ Proposition 1.3} (a) $\uqg = U^-.U^0.U^+.$

(b) $\uqgh = \hat U^-.\hat U^0.\hat U^+.$ \qed
\endproclaim
See [5] or [10] for details.

A representation $W$ of $\uqg$ is said to be of type 1 if it is the direct sum
of its weight spaces
$$W_{\lambda} =\{w\in W\mid k_i.w = q_i^{\lambda(i)}w\},\ \ \ \ \ (\lambda\in
P).$$
If $W_\lambda\ne 0$, then $\lambda$ is a weight of $W$. A vector $w\in
W_{\lambda}$ is a highest weight vector if $x_i^+.w =0$ for all $i\in I$, and
$W$ is a highest weight representation with highest weight $\lambda$ if
$W=\uqg.w$ for some highest weight vector $w\in W_\lambda$.

It is known (see [5] or [10], for example) that every finite--dimensional
irreducible representation of $\uqg$ of type 1 is  highest  weight. Moreover,
assigning to such a representation its highest weight defines a bijection
between the set of isomorphism classes of finite--dimensional irreducible type
1 representations of $\uqg$ and $P^+$; the irreducible type 1 representation of
$\uqg$ of highest weight $\lambda\in P^+$ is denoted by $V(\lambda)$. Finally,
every finite--dimensional representation $W$ of $\uqg$ is completely reducible:
if $W$ is of type 1, then
$$W\simeq\bigoplus_{\lambda\in P^+}V(\lambda)^{\op m_{\lambda}(W)}$$
for some uniquely determined multiplicities $m_{\lambda}(W)\in\Bbb N$.
It is convenient to introduce the following notation: for $\mu\in P^+$, let
$$W_{\mu}^+ =\{w\in W_\mu: x_{i,0}^+.v =0 \ \ \text {for all $i\in I$}\}.$$
Then, $m_\mu(W) ={\text dim}(W_\mu^+)$.

A representation $V$ of $\uqgh$ is of type 1 if $c^{1/2}$ acts as the identity
on $V$, and if $V$ is of type 1 as a representation of $\uqg$. A vector $v\in
V$ is a highest weight vector if
$$x_{i,r}^+.v=0,\ \ \phi_{i,r}^{{}\pm{}}.v=\Phi_{i,r}^{{}\pm{}}v,\ \ \ c^{1/2}.
v =v,$$
for some complex numbers $\Phi_{i,r}^{{}\pm{}}$. A type 1 representation $V$ is
a highest weight representation if $V=\uqgh.v$, for some highest weight vector
$v$, and the pair of $(I\times\Bbb Z)$-tuples $(\Phi_{i,r}^{{}\pm{}})_{i\in
I,r\in\Bbb Z}$ is its highest weight. Note that $\Phi_{i,r}^+=0$ (resp.
$\Phi_{i,r}^-=0$) if $r<0$ (resp. if $r>0$), and that
$\Phi_{i,0}^+\Phi_{i,0}^-=1$. (In [5], highest weight representations of
$\uqgh$ are called `pseudo-highest weight'.) Lowest weight representations are
defined similarly.

If $\lambda\in P^+$, let ${\Cal P}^\lambda$ be the set of all $I$-tuples
$(P_i)_{i\in I}$ of polynomials $P_i\in\Bbb C[u]$, with constant term 1, such
that $deg(P_i)=\lambda(i)$ for all $i\in I$. Set ${\Cal P}=\cup_{\lambda\in
P^+}{\Cal P}^\lambda$.

\proclaim{Theorem 1.4} (a) Every finite-dimensional irreducible representation
of $\uqgh$ can be obtained from a type 1 representation by twisting with an
automorphism of $\uqgh$.

(b) Every finite-dimensional irreducible representation of $\uqgh$ of type 1 is
both  highest and lowest  weight.

(c) Let $V$ be a finite-dimensional irreducible representation of $\uqgh$ of
type 1 and highest weight $(\Phi_{i,r}^{{}\pm{}})_{i\in I,r\in\Bbb Z}$. Then,
there exists $\bp=(P_i)_{i\in I}\in\calp$ such that
$$\sum_{r=0}^\infty\Phi_{i,r}^+u^r=q_i^{deg(P_i)}\frac{P_i(q_i^{-2}u)}{P_i(u)}=\sum_{r=0}^\infty\Phi_{i,r}^-u^{-r},$$
in the sense that the left- and right-hand terms are the Laurent expansions of
the middle term about $0$ and $\infty$, respectively. Assigning to $V$ the
$I$-tuple $\bp$ defines a bijection between the set of isomorphism classes of
finite-dimensional irreducible representations of $\uqgh$ of type 1 and
$\calp$. We denote by $V(\bold P)$ the irreducible representation associated to
$\bold P$.

(d) Let $\bp$, $\bq\in\calp$ be as above, and let $v_{\bp}$ and $v_\bq$ be
highest weight vectors of $V(\bp)$ and $V(\bq)$, respectively. Then, in
$V(\bp)\ot V(\bq)$,
$$x_{i,r}^+.(v_\bp\ot v_\bq)=0,\ \ \phi_{i,r}^{{}\pm{}}.(v_\bp\ot
v_{\bq})=\Psi_{i,r}^{{}\pm{}}(v_\bp\ot v_\bq),$$
where the complex numbers $\Psi_{i,r}^{{}\pm{}}$ are related to the polynomials
$P_iQ_i$ as the $\Phi_{i,r}^{{}\pm{}}$ are related to the $P_i$ in part (c). In
particular, if $\bp\ot\bq$ denotes the $I$-tuple $(P_iQ_i)_{i\in I}$, then
$V(\bp\ot\bq)$ is isomorphic to a quotient of the subrepresentation of
$V(\bp)\ot V(\bq)$ generated by $v_{\bp}\ot v_{\bq}$.

(e) If $\bp=(P_i)_{i\in I}\in\calp$, $a\in\Bbb C^\times$, and if
$\tau_a^*(V(\bp))$ denotes the pull-back of $V(\bp)$ by the automorphism
$\tau_a$, we have
$$\tau_a^*(V(\bp))\cong V(\bp^a)$$
as representations of $\uqgh$, where $\bp^a=(P_i^a)_{i\in I}$ and
$$P_i^a(u)=P_i(au).\ \ \ \ \qed$$\endproclaim

See [5] and [7] for further details. If the highest weight
$(\Phi_{i,r}^{{}\pm{}})_{i\in I,r\in\Bbb Z}$ of $V$ is given by an $I$-tuple
$\bp$ as in part (c), we shall often abuse notation by saying that $V$ has
highest weight $\bp$.

If $a\in\Bbb C^{\times}$, $i\in I$, we denote the irreducible representation of
$\uqgh$ with defining polynomials
$$P_j =\cases
1 & \text{if $j\ne i$,}\\
1-a^{-1}u & \text{if $j=i$}\endcases$$
by $V(\lambda_i, a)$, and denote the highest (resp. lowest) weight vector in
this representation by $v_{\lambda_i}$ (resp. $v_{-\lambda_i}$).

For $i\in I$,  the subalgebra of $\uqgh$ generated by the elements
$x_{i,r}^{{}\pm{}}$ ($r\in\Bbb Z$), $k_i^{{}\pm 1}$ ($i\in I$), and $h_{i,r}$
($i\in I$, $r\in\Bbb Z\backslash\{0\}$) is isomorphic to $U_{q_i}(\hat{sl}_2)$;
we denote this subalgebra by $U_q(\hat\ung_{(i)})$. The subalgebra
$U_{q}(\ung_{(i)})$ is defined similarly. Let $\mu_{(i)}$ be the restriction of
$\mu$ to $\{i\}$. The following lemma was proved in [6].
\proclaim{Lemma 1.5} Let $M$ be any highest weight representation of $\uqgh$
with highest weight $P$ and highest weight vector $m$.

\noindent (i) For $i=1,2$, $M_{(i)} =  U_q(\hat{\ung}_{(i)}).m$ is a highest
weight representation of $U_q(\hat{\ung}_{(i)})$ with highest weight $P_i$ and
$$m_{\mu}(M) = m_{\mu_{(i)}}(M_{(i)}).$$
(ii) Let $N$ be another highest weight representation of $\uqgh$ with highest
weight $\bq$ and assume that $\lambda$ is the highest weight of $M\ot N$ (i.e.
$\lambda(i) =\text{deg}(P_i) +\text{deg}(Q_i)$ for $i=1,2$). Then, for $i=1,2$
and $r\in\Bbb Z_+$, we have
$$m_{\lambda-r\alpha_i}(M\ot N) = m_{ \lambda_{(i)}-r\alpha_i}(M_{(i)}\ot
N_{(i)}).\ \ \ \ \qed$$

\endproclaim

\vskip 36pt

\noindent {\bf 2 Minimal affinizations}
\vskip12pt\noindent
Following [2], we say that a finite-dimensional irreducible representation $V$
of $\uqgh$ is an affinization of $\lambda\in P^+$ if $V\cong V(\bp)$ as a
representation of $\uqgh$, for some $\bp\in{\Cal P}^\lambda$. Two affinizations
of $\lambda$ are equivalent if they are isomorphic as representations of
$\uqg$; we denote by $[V]$ the equivalence class of $V$. Let ${\Cal Q}^\lambda$
be the set of equivalence classes of affinizations of $\lambda$.

The following result is proved in [2].
\proclaim{Proposition 2.1} If $\lambda\in P^+$ and $[V]$,
$[W]\in\Cal{Q}^{\lambda}$, we write $[V]\preceq [W]$ iff, for all $\mu\in P^+$,
either,

(i) $m_{\mu}(V)\le m_{\mu}(W)$, or

(ii) there exists $\nu>\mu$ with $m_{\nu}(V) < m_{\nu}(W)$.

\noindent Then, ${}\preceq{}$ is a partial order on ${\Cal
Q}^\lambda$.\qed\endproclaim

An affinization $V$ of $\lambda$ is minimal if $[V]$ is a minimal element of
$\Cal{Q}^{\lambda}$ for the partial order $\preceq$, i.e. if
$[W]\in\Cal{Q}^{\lambda}$ and $[W]\preceq [V]$ implies that $[V]=[W]$.
It is proved in [2] that ${\Cal Q}^\lambda$ is a finite set, so minimal
affinizations certainly exist.

A necessary condition for minimality was obtained in [2]. To state this result,
we recall that  the set of complex numbers $\{aq^{-r+1}, aq^{-r+3}, \ldots ,
aq^{r -1}\}$ is called the $q$--segment of length $r\in\Bbb N$  and centre
$a\in\Bbb C^{\times}$.

\proclaim{Proposition 2.2} Let $\lambda\in P^+$, let $\bp=(P_i)_{i\in
I}\in\calp^\lambda$, and assume that $V(\bp)$ is a minimal affinization of
$\lambda$. Then, for all $i\in I$, the roots of $P_i$ form a $q_i$-segment of
length $\lambda(i)$.\qed\endproclaim

Note that it follows from 1.4(e) and 2.2 that, if $i\in I$ and $r\in\Bbb N$,
the weight $r\lambda_i$ has a unique affinization, up to equivalence.

{\it For the rest of this paper we assume that $\ung$ is of type $B_2$.}
In this case, the defining polynomials of the minimal affinizations were
determined in [2]:
\proclaim{Theorem 2.3}
Let $\lambda\in P^+$ and $\bp\in\calp^{\lambda}$. Then, $V(\bp)$ is a minimal
affinization of $\lambda$ iff the following conditions are satisfied:

\noindent (a) for each $i=1,2$, either $P_i =1$ or the roots of $P_i$ form a
$q_i$--segment  of length $\lambda(i)$ and centre $a_i$ (say);

\noindent (b) if $P_1\ne 1$ and $P_2\ne 1$, then
$$\frac{a_1}{a_2} =q^{2\lambda(1)+\lambda(2)+1}\ \ \ {\text{or}}\ \
q^{-(2\lambda(1)+\lambda(2)+3)}.$$
Any two minimal affinizations of $\lambda$ are equivalent. Finally, if $V(\bp)$
is a minimal affinization of $\lambda$ and $r\in\Bbb Z_+\backslash\{0\}$, we
have
$$m_{\lambda -r\alpha_1}(V(\bp)) = m_{\lambda-r\alpha_2}(V(\bp))
=m_{\lambda-\alpha_1-\alpha_2}(V(\bp)) =0.\qed$$\endproclaim
Our concern in this paper is the structure of a minimal affinization $V(\bp)$
as a representation of $\uqg$.  Our main result is:
\proclaim{Theorem 2.4} Let $\lambda\in P^+$ and let $V(\bp)$ be a minimal
affinization of $\lambda$. Then, as a representation of $\uqg$,
$$V(\bp)\cong \bigoplus_{r=0}^{{\text{int}}(\frac12\lambda(2))}V(\lambda
-2r\lambda_2).$$\endproclaim
Here, for any real number $b$, $\text{int}(b)$ is the greatest integer less
than or equal to $b$.

The proof of Theorem 2.4 is by induction on $\lambda(2)$. The first part of the
following proposition begins the induction.
\proclaim{Proposition 2.5}

\noindent (a) For any $r\in\Bbb N$, the minimal affinization of $r\lambda_1$ is
irreducible as a $\uqg$--module.

\noindent (b) The minimal affinization of $\lambda_2$  is irreducible as a
representation of $\uqg$.
\endproclaim
\demo{Proof} (a)  Let $\bp\in\calp^{r\lambda_1}$ be such that $V(\bp)$ is a
minimal affinization of $r\lambda_1$.
The element  $x_0^+.\vbp$ has weight $r_1\lambda_1-\alpha_1 -2\alpha_2$. This
weight  is Weyl group conjugate to $r_1\lambda_1-\alpha_1\in P^+$. Hence, if
$m_\nu(V(\bp)>0$ and $x_0^+.\vbp$ has a non-zero component in a
$\uqg$-subrepresentation of $V(\bp)$ of highest weight $\nu$, then
$\nu=r_1\lambda_1$ or $r_1\lambda_1-\alpha_1$. But,
$m_{r_1\lambda_1-\alpha_1}\!(V(\bp))=0$ by 2.2 and so $x_0^+.
\vbp\in\uqg.\vbp\cong V(r_1\lambda_1)$.  It follows that $x_0^+$ preserves
$V(r\lambda_1)$. Working with a lowest weight vector of $V(\bp)$, one proves
similarly that $x_0^{-}$ preserves $\uqg.\vbp$. Hence,  $\uqg.\vbp$ is a
$\uqgh$-subrepresentation of $V(\bp)$, hence is equal to $V(\bp)$, and so
$V(\bp)\cong V(r\lambda_1)$ as representations of $\uqg$.

(b) This is obvious, since there is no $\mu\in P^+$ such that $\mu
<\lambda_2$.\qed\enddemo

We conclude this section with the following result on the dual of
$V(\lambda_2,a)$.

If $V$ is any representation of $\uqgh$, its left dual $^tV$ is the
representation of $\uqgh$ on the vector space dual of $V$ given by
$$<a.f, v>\ \  = \ \ < f, S(a). v>,\ \ \ \ \ \ \ (a\in\uqgh , v\in V, f\in
{}^tV) $$
where $S$ is the antipode of $\uqgh$ and $< , >$ is the natural pairing between
$V$ and its dual. The right dual $V^t$ is defined in the same way, replacing
$S$ by $S^{-1}$. Left and right duals of representations of $\uqg$ are defined
similarly. Clearly the (left or right) dual of an irreducible representation is
again irreducible. In fact, it is well known that, for any $\lambda\in P^+$,
$$^tV(\lambda)\cong V(\lambda)^t\cong V(-w_0\lambda),$$
where $w_0$ is the longest element of the Weyl group of $\ung$.

\proclaim{Lemma 2.6} (i) For any $a\in\Bbb C^{\times}$,
$$V(\lambda_2, a)^t\cong V(\lambda_2, aq^6), \ \ \ \ ^tV(\lambda_2, a)\cong
V(\lambda_2, aq^{-6}).$$

\noindent (ii) For any $a,b\in\Bbb C^{\times}$,
$$\text{dim}((V(\lambda_2,a)\ot V(\lambda_2,b))_0^+)= 1.$$
Moreover, if $0\ne v_0\in (V(\lambda_2,a)\ot V(\lambda_2,b))_0^+$ and $a/b\ne
q^{{}\pm 6}$, then $x_0^{{}\pm{}}.v_0$ is a non--zero multiple of
$v_{{}\mp\lambda_2}\ot v_{{}\mp\lambda_2}$.

\endproclaim
\demo{Proof} (i)  Since, for any representation $V$ of $\uqgh$, the canonical
isomorphism of vector spaces $^tV^t\to V$ is an isomorphism of representations,
it suffices to prove the first formula. Since $V(\lambda_2)$ is a self-dual
representation of $\uqg$, we have {\it a priori} that $V(\lambda_2,a)^t\cong
V(\lambda_2,b)$ for some $b\in\Bbb C^{\times}$.

Fix $v_{-\lambda_2} =x_2^-x_1^-x_2^-.v_{\lambda_2}$. Then, $v_{-\lambda_2}$ is
a non--zero element of $V(\lambda_2,a)_{-\lambda_2}$ and, for weight reasons,
$$x_0^+.v_{\lambda_2} = Av_{-\lambda_{2}}$$ for some $A\in\Bbb C$.
  Let $0\ne v_{\lambda_2}^t\in V(\lambda_2,a)^t_{\lambda_2}$. Then
$<v_{\lambda_2}^t, w>\ =0$ if $w\notin V(\lambda_2,a_2)_{-\lambda_2}$.
Normalize $v_{\lambda_2}^t$ so that
$$ <v_{\lambda_2}^t, v_{-\lambda_2}> =1,$$
and let $v_{-\lambda_2}^t =x_2^-x_1^-x_2^-.v_{\lambda_2}^t$. Again, for weight
reasons, one has
$$x_0^+.v_{\lambda_2}^t = Bv_{-\lambda_2}^t.$$
for some  $B\in \Bbb C$. Moreover, from the formula for $x_0^+$ in 1.2, it is
clear that
$$A =a^{-1}c,\ \ \  B= b^{-1}c,$$
where $c\in\Bbb C^{\times} $ depends only on $q$, and not on $a$ or $b$. Thus,
$A/B = b/a$. But $A/B$ may be computed as follows:
$$<x_0^+.v_{\lambda_2}^t, v_{\lambda_2}>\ \ =\ \  < v_{\lambda_2}^t,
S^{-1}(x_0^+).v_{\lambda_2}> \ = \ < v_{\lambda_2}^t,
-k_0^{-1}x_0^+.v_{\lambda_2}>.$$
Hence,
$$B<x_2^-x_1^-x_2^-.v_{\lambda_2}^t, v_{\lambda_2}>\  = -q^{-2}A.$$
Since
$$S(x_2^-x_1^-x_2^-).v_{\lambda_2} = q^4x_2^-x_1^-x_2^-.v_2 =
q^4v_{-\lambda_2},$$
we find that $A/B =q^6$, and part (i) is proved.

(ii) Since $V(\lambda_2)$ is a self--dual representation of $\uqg$, it follows
from 2.4 that
$$\text{dim}((V(\lambda_2,a)\ot V(\lambda_2,b))_0^+)= 1.$$
If $x_0^{{}\pm{}}.v_0  =0$, then $\Bbb C.v_0$ is a $\uqgh$--subrepresentation
of the tensor product, and hence by (i) we have $a/b =q^{{}\pm 6}$.
\qed\enddemo

\vskip36pt

\noindent{\bf 3 A first reduction}
\vskip12pt\noindent {\it For the remainder of this paper}, we assume that
$\lambda\in P^+$, $\lambda(2)\ge 1$ and that 2.4 is known for
$\lambda-\lambda_2$. We shall also assume that $\lambda(1) \ge 1$. The proof
when $\lambda(1) =0$ is  similar and easier.

We also {\it fix for the rest of the paper} an element $\bp=(P_i)_{i\in
I}\in\calp^{\lambda}$ such that the roots of $P_i$ form a string with centre
$a_i$ and length $\lambda(i)$, $i=1,2$, and such that
$$\frac{a_1}{a_2} =q^{-(2\lambda(1)+\lambda(2)+3)}.$$
Define an element $\bq\in \calp^{\lambda-\lambda_2}$ by
$$Q_1 =P_1,\ \ \ Q_2=\prod_{i=1}^{\lambda(2)-1}(1-a_2^{-1}q^{-(\lambda(2)
-2i-1)}u).$$
By 2.3, $V(\bp)$ and  $V(\bq)$ are  minimal affinizations of $\lambda$ and
$\lambda-\lambda_2$, respectively. In particular, 2.4 is known for $V(\bq)$.

The following is the main result of this section:
\proclaim{Proposition 3.1}  Let $\lambda, \mu\in P^+$ and let $V(\bp)$ be a
minimal affinization of $\lambda$ as above. Then:

(i) $m_{\mu}(V(\bp))\le 1$ if $\mu$ is of the form $\lambda-r\theta-\alpha_2$
or $\lambda-r\theta-\alpha_1-\alpha_2$ for some $r\in\Bbb N$.

(ii) $ m_{\mu}(V(\bp))\le 2$ if $\mu$ is of the form $\lambda-r\theta$ for some
$r\in\Bbb N$.

(iii) $m_{\mu}(V(\bp)) =0$ if $\mu$ is not of the form $\lambda-r\theta$,
$\lambda-r\theta-\alpha_2$ or $\lambda-r\theta-\alpha_1-\alpha_2$ for some
$r\in\Bbb N$.

(iv) $m_{\lambda-r\theta}(V(\bp))\ge 1$ for $0\le
r\le{\text{int}}(\frac12\lambda(2))$.

\endproclaim

We deduce this from the next two results.
\proclaim{Lemma 3.2 } For any $\lambda\in P^+$,
$$V(\lambda)\ot V(\lambda_2)\cong V(\lambda+\lambda_2)\op
V(\lambda+\lambda_2-\alpha_2)\op V(\lambda+\lambda_2-\alpha_1-\alpha_2)\op
V(\lambda+\lambda_2-\theta).$$
\endproclaim
\demo{ Proof} By 1.4(c), it suffices to prove the analogous classical result.
We leave this to the reader.\qed\enddemo

\proclaim{Proposition 3.3} Let $\lambda\in P^+$, $\bp\in\calp^{\lambda}$,
$\bq\in\calp^{\lambda-\lambda_2}$ be as defined above.

\noindent (i) $V(\lambda_2,a_2q^{\lambda(2)-1})\ot V(\bq)$  is generated as a
representation of $\uqgh$ by the tensor product of the highest weight vectors.
In particular, $V(\bp)$ is isomorphic to a quotient of
$V(\lambda_2,a_2q^{\lambda(2)-1})\ot V(\bq)$.

\noindent (ii) Let $\bp_{(1)} = (P_1, 1)$. Then, there exists a surjective
homomorphism of representations of $\uqgh$
$$\pi:V(\lambda_2, a_2q^{\lambda(2)-1})\ot V(\lambda_2,
a_2q^{\lambda(2)-3})\ot\cdots\ot V(\lambda_2,a_2q^{-\lambda(2)+1})\ot
V(\bp_{(1)})\to V(\bp)$$
such that $\pi(v_{\lambda_2}^{\ot \lambda(2)}\ot v_{\bp_{(1)}}) =\vbp$.
\endproclaim
We assume 3.3 for the moment and give the
\demo {Proof of 3.1}   Parts (i), (ii) and (iii)  are easy consequences of
2.4(ii), 3.2 and 3.3(i), since 2.4 is known for $V(\bq)$.

To prove (iv), we can assume that $\lambda(2) \ge 2$, since otherwise there is
nothing to prove.  Notice that, by 2.5, we can (and do) choose elements  $0\ne
w_s\in (V(\lambda_2, a_2q^{\lambda(2)-4s+3})\ot V(\lambda_2,
a_2q^{\lambda(2)-4s+1}))_0^+$ such that
$$x_0^-.w_s =v_{\lambda_2}\ot v_{\lambda_2}.$$
For $1\le r\le\text{int}(\frac12 \lambda(2))$, consider the element $w =w_1\ot
w_2\ot\cdots w_r\ot v_{\lambda_2}^{\ot\lambda(2)-2r}\ot v_{\bold P_{(1)}}\in
V(\lambda_2, a_2q^{\lambda(2)-1})\ot V(\lambda_2,
a_2q^{\lambda(2)-3})\ot\cdots\ot V(\lambda_2,a_2q^{-\lambda(2)+1})\ot
V(\bp_{(1)})$. Clearly, $x_{i,0}^+.w =0$ for $i=1,2$, and an easy computation
shows that
$$(x_0^-)^r.w = q^{r(r-1)}[r]_{q^2}v_{\lambda_2}^{\ot \lambda(2)}\ot
v_{\bp_{(1)}}.$$
Hence, $\pi((x_0^-)^r.w)\ne 0$ and so $\pi(w)$ is a non--zero element of
$V(\bp)_{\lambda-r\theta}^+$. This proves 3.1(iv).
\qed\enddemo

\demo{Proof of 3.3} Assuming 3.3(i) we give the proof of 3.3(ii).
The proof is by induction on $\lambda(2)$. The case $\lambda(2) =1$ is just
3.3(i). So if $\lambda(2) >1$, by the induction hypothesis applied to
$\lambda-\lambda_2$,  we have a surjective homomorphism of representations of
$\uqgh$
$$\pi': V(\lambda_2, a_2q^{\lambda(2)-3})\ot\cdots\ot V(\lambda_2,
a_2q^{-\lambda(2)+1})\ot V(\bold{P}_{(1)})\to V(\bq).$$
Consider $$ {\text{id}}\ot\pi':
 V(\lambda_2, a_2q^{\lambda(2)-1})\ot\cdots\ot V(\lambda_2,
a_2q^{-\lambda(2)+1})\ot V(\bold{P}_{(1)})
\to V(\lambda_2, a_2q^{\lambda(2)-1})\ot V(\bq).$$
By 3.3(i), the right-hand side has $V(\bp)$ as a quotient and so we get  the
required surjective homomorphism
$$\pi: V(\lambda_2, a_2q^{\lambda(2)-1})\ot\cdots\ot V(\lambda_2,
a_2q^{-\lambda(2)+1})\ot V(\bold{P_{(1)}})\to V(\bp).$$

We now prove 3.3(i). Let $M =\uqgh.(v_{\lambda_2}\ot v_{\bq})$. We first show
that it suffices to prove
$$m_{\mu}(M) =m_{\mu}(V(\lambda_2, a_2q^{\lambda(2)-1})\ot V(\bq))\ \ \ \
{\text{for}} \ \ \mu > \lambda-\theta .\tag3$$
To see this, assume that $M$ is a proper subrepresentation of the tensor
product and let $N$ be the corresponding quotient. It follows from 3.2 and (3)
that
$$m_{\mu}(N) = 0\ \ \ \ {\text{unless}}\ \ \mu\le\lambda-\theta .$$
On the other hand,
 dualizing the projection map
$$V(\lambda_2, a_2q^{\lambda(2)-1})\ot V(\bq)\to N$$
we get a non--zero (hence injective) homomorphism of representations of $\uqgh$
$$V(\bq)\to V(\lambda_2, a_2q^{\lambda(2)})^t\ot N.$$
It follows that
$$m_{\lambda-\lambda_2}(V(\lambda_2)\ot N) \ge 1,$$
and hence by 3.2 that $m_{\lambda-\theta}(N)> 0$.

Note that the preceding argument proves that, if $M'$ is any
$\uqgh$--subrepresentation of $V(\lambda_2, a_2q^{\lambda(2)-1})\ot V(\bq)$
containing $M$, and if $N'$ is the corresponding quotient of the tensor
product, then $m_{\lambda-\theta}(N') >0$.
In particular, any irreducible quotient $N'$ of $N$ must have
$m_{\lambda-\theta}(N') >0$. Taking $N'$ to be an  affinization $V(\bold{R})$,
say, of $\lambda-\theta$, we have a surjective map of $\uqgh$--representations
$$V(\lambda_2, a_2q^{\lambda(2)})\ot V(\bq)\to V(\bold{R}),$$
and hence, dualizing, an injective map
$$V(\bq)\to V(\lambda_2 , a_2q^{\lambda(2)-1})^t\ot V(\bold{R}) = V(\lambda_2,
a_2q^{\lambda(2)+5})\ot V(\bold{R}),$$
by 2.5.  The highest weight vector in $V(\bq)$ must map to (a non--zero
multiple of) the tensor product of the highest weight vectors on the right-hand
side. But this is impossible, since $a_2q^{\lambda(2)+5}$ is not a root of
$Q_2$. Hence, $N =0$ and part (i) is proved.

We now prove (3). The statement is obviously true for $\mu =\lambda$. For
$\mu=\lambda-\alpha_1$, the statement follows from 3.2 and the fact that 2.4 is
known for $V(\bq)$. For $\mu=\lambda-\alpha_2$, notice that, by 1.5, it
suffices to prove the result for $U_q(\hat{sl}_2)$. But this follows from
4.9(a) of [3] since
$$V(\lambda_2, a_2q^{\lambda(2)-1})_{(2)}\ot V(\bq)_{(2)}\cong V(1,
a_2q^{\lambda(2)-1})\ot V(\lambda(2) -1, a_2q^{-1})$$
as representations of $U_q(\hat{\ung}_{(2)})$.

Finally, we must prove (3) for $\lambda-\alpha_1-\alpha_2$.
For this, it obviously suffices to prove that
$$(V(\lambda_2, a_2q^{\lambda(2)-1})\ot
V(\bq))_{\lambda-\alpha_1-\alpha_2}\subseteq M.$$
The left-hand side is spanned by
$$\{x_1^-x_2^-.v_{\lambda_2}\ot\vbq, v_{\lambda_2}\ot x_1^-x_2^-.\vbq,
v_{\lambda_2}\ot x_2^-x_1^-.\vbq, x_2^-.v_{\lambda_2}\ot x_1^-.\vbq\}.$$
Now, since $m_{\lambda}(M)$ and $m_{\lambda-\alpha_2}(M)$ are both strictly
positive, $M$ contains $x_2^-.v_{\lambda_2}\ot\vbq$ and $v_{\lambda_2}\ot
x_2^-.\vbq$ (since $M$ contains two linear combinations of these vectors which
are not scalar multiples of each other). Also, $M$ contains
$$v_{\lambda_2}\ot x_1^-.\vbq =x_1^-.(v_{\lambda_2}\ot\vbq).$$
It follows that $M$ contains the three vectors
$$x_1^-.(x_2^-.v_{\lambda_2}\ot\vbq),\ \  x_1^-.(v_{\lambda_2}\ot x_2^-.\vbq),\
\ \ x_2^-.(v_{\lambda_2}\ot x_1^-.\vbq),$$
i.e. that $M$ contains the three vectors
$$\aligned
v_{\lambda_2}&\ot x_1^-x_2^-.\vbq, \\
x_1^-x_2^-.v_{\lambda_2}\ot\vbq &+ q^{-2} x_2^-.v_{\lambda_2}\ot x_1^-.\vbq, \\
x_2^-.v_{\lambda_2}\ot x_1^-.\vbq &+ q^{-1} v_{\lambda_2}\ot x_2^-x_1^-.\vbq
.\endaligned\tag4$$
Since these vectors are obviously linearly independent, it suffices to prove
that
$$ [x_2^+, x_0^+].(v_{\lambda_2}\ot\vbq) \tag5$$
is linearly independent of the vectors in (4).

To compute the vector in (5), we need the following formulas:
$$\align x_{1,1}^-.\vbq& =a_1^{-1}q^{2\lambda(1)-2}x_1^-.\vbq,\tag6\\
x_{1,1}^-x_2^-.\vbq&=a_1^{-1}q^{2\lambda(1)-2}x_1^-x_2^-.\vbq.\tag7\endalign$$

By the formula for the isomorphism $f$ in 1.2,
$$[x_2^+,x_0^+] =c[x_2^-,x_{1,1}^-]_q(k_1k_2)^{-1},$$
where $c\in\Bbb C^{\times}$, and for any $x,y\in\uqgh$, we define
$$[x,y]_q =qxy-q^{-1}yx.$$
Using this, we find that
$$\align
[x_2^+, x_0^+].(v_{\lambda_2}\ot\vbq)
&=cq^{-(2\lambda(1)+\lambda(2))}(q^{-1}[x_2^-,x_{1,1}^-]_q.v_{\lambda_2}\ot\vbq
+v_{\lambda_2}\ot [x_2^-,x_{1,1}^-]_q.\vbq)\\
&=cq^{-(2\lambda(1)+\lambda(2))}(-a_2^{-1}q^{-\lambda(2)+1}x_1^-x_2^-.
v_{\lambda_2}\ot\vbq\\
&\qquad\qquad +v_{\lambda_2}\ot (a_1^{-1}q^{2\lambda(1)-1}x_2^-x_1^-.\vbq
-a_2^{-1}q^{\lambda(2) -3}x_1^-x_2^-.\vbq)).
\endalign$$
An easy computation shows that this is linearly dependent on the vectors in (5)
iff
$$\frac{a_1}{a_2} = q^{2\lambda(1) +\lambda(2)+2},$$
contradicting our assumption that $a_1/a_2 =q^{-(2\lambda(1)+\lambda(2)+2)}$.

The proof of (6) is easy since we know from 2.2 that $x_{1,1}^-.\vbq$ must be a
scalar multiple of $x_{1}^-.\vbq$.  As for (7), observe that by 2.2 again we
know {\it a priori} that
$$x_{1,1}^-x_2^-.\vbq = Ax_1^-x_2^-.\vbq +B x_2^-x_1^-.\vbq$$
for some $A, B\in\Bbb C$. Applying $x_{1}^+$ and $x_{2}^+$ to both sides of
gives the pair of equations
$$\align
A[\lambda(2)-1]_q
+B[\lambda(2)]_q&=[\lambda(2)-1]_qa_1^{-1}q^{2\lambda(1)-2},\\
A[\lambda(1)+1]_{q^2}+B[\lambda(1)]_{q^2}&=q^{2\lambda(1)a_1^{-1}}[\lambda(1)]_{q^2}+a_2^{-1}q^{(2\lambda(1)+\lambda(2)+1)}.
\endalign$$
Using $a_1/a_2 = q^{-(2\lambda(1)+\lambda(2)+3)}$, we find that the unique
solution is $A =a_1^{-1}q^{2\lambda(1)-2}$, $B=0$.
\qed\enddemo

\vskip 36pt
\noindent{\bf 4. Completion of the proof of Theorem 2.4}
\vskip12pt

In view of 3.3, to complete the proof of 2.4, it suffices to establish
\proclaim{Proposition 4.1} Let $\lambda\in P^+$ and let $V(\bp)$ be a minimal
affinization of $\lambda$. Then:

(i) $m_{\lambda-r\theta}(V(\bp)) =1$ if $0\le
r\le{\text{int}}(\frac12\lambda(2))$.

(ii) $m_{\mu}(V(\bp)) =0$ if $\mu$ is of the form $\lambda-r\theta-\alpha_2$,
for some $r\in\Bbb N$.

(iii) $m_{\mu}(V(\bp))=0$ if $\mu$ is of the form
$\lambda-r\theta-\alpha_1-\alpha_2$ for some $r\in\Bbb N$.
\endproclaim

We need three lemmas.

\proclaim{Lemma 4.2}  Suppose that there exists  $0\ne v\in V(\bp)_{\mu}^+$
such that $$x_{1,1}^+.v =x_{1,-1}^+.v =0$$ (resp. $x_{2,1}^+.v =x_{2,-1}^+.v
=0$). Assume that $m_{\mu+\alpha_i}(V(\bp)) =0$ for $i=1,2$. Then,
$\lambda=\mu$.
\endproclaim
\demo{Proof} We prove, by induction on $k\in \Bbb N$, that
$$x_{i, \pm k}^+. v =0 \ \ {\text{for all}}\ \  i=1,2.\tag8$$
 It is easy to see using the relations in 1.2 that the $k_j$ and $h_{j,s}$
preserve the finite--dimensional space
$$V(\bp)_{\mu}^{++} =\{w\in V(\bp)_{\mu}: x_{i,k}^+.w =0 \ \text{forall}\ \
i\in I, k\in\Bbb Z\}.$$
It follows that there exists a $\uqgh$--highest weight vector in $V(\bp)_\mu$,
which is possible only if $\lambda=\mu$.

It is obvious that (8) holds when $k=0$. We assume that it holds for $k$ and
prove it for $k+1$. Using (1), we find that
$$x_{j,0}^+x_{i, \pm (k+1)}^+\in\uqgh x_{j,0}^++\uqgh x_{j,\pm 1}^++\uqgh
x_{i,\pm k}^+,$$
and hence by the induction hypotheses we see that $x_{i,\pm k+1}^+.v\in
V(\bp)_{\mu+\alpha_i}^+$. Since  $m_{\mu+\alpha_i}(V(\bp)) =0$ by assumption,
this forces $x_{i,\pm k+1}^+.v =0$, establishing (8) for $k+1$.\qed\enddemo

\proclaim{Lemma 4.3}Let $0\ne v\in V(\bp)_{\mu}$ be such that $x_{1,s'}^+.v =0$
if $0\le s' <s$ or if $s<s'\le 0$. Then

\noindent (i) $(x_{2,0}^+)^3x_{1,s}^+.v =0$,

\noindent (ii) $x_{1,0}^+x_{2,0}^+x_{1,s}^+. v =0,$

\noindent (iii) $x_{1,0}^+(x_{2,0}^+)^2x_{1,s}^+.v \in
V(\bp)_{\mu+2\alpha_1+2\alpha_2}^+$.
\endproclaim
\demo{Proof} Using relation (2) we find that
$$ (x_{2,0}^+)^3x_{1,\pm s}^+\in \uqgh.x_{2,0}^+.\tag9$$
Part (i) is now immediate.

For (ii), it suffices to notice that (1) and (2) together give
$$x_{1,0}^+x_{2,0}^+x_{1,\pm s}^+\in \uqgh x_{2,0}^+ +\sum_{0\le s'<s}\uqgh
x_{1,\pm s'}^+\tag10$$
if $s>0$.

For (iii), we use  the following consequences of (1) and  (2):
$$\align (x_{1,0}^+)^2x_{2,0}^+&\in\uqgh x_{1,0}^+,\tag11\\
x_{2,0}^+x_{1,0}^+(x_{2,0}^+)^2&\in\uqgh(x_{2,0}^+)^3 +\uqgh x_{1,0}^+ +\uqgh
x_{1,0}^+x_{2,0}^+.\tag12\endalign$$
The result now follows from parts (i) and (ii).\ \qed\enddemo

\proclaim{Lemma 4.4}  Let $\mu\in P^+$ be  such that
$$m_{\mu+\eta}(V(\bp)) =0 \ \ {\text{if}}\ \  \eta\ne s\theta , s\in\Bbb Z_+
.\tag13$$
Then, $(x_{2,0}^+)^2x_{1,\pm 1}^+$ maps $V(\bp)_{\mu}^+$ to
$V(\bp)_{\mu+\theta}^+$.
Further, if $v\in V(\bp)_{\mu}^+$ is such that $x_{1,\pm 1}^+.v\ne 0$, then
$$(x_{2,0}^+)^2x_{1,\pm 1}^+\ne 0 .$$
\endproclaim

 \demo{Proof} It is clear for weight reasons that $(x_{2,0}^+)^2x_{1,\pm 1}^+$
maps $V(\bp)_{\mu}^+$ to $V(\bp)_{\mu+\theta}$. Thus, it suffices to prove that
$$x_{i,0}^+(x_{2,0}^+)^2x_{1,\pm 1}^+. v = 0\ \ \text{if} \ \ v\in
V(\bp)_{\mu}^+.$$
For $i=2$, this is just 4.3(i). For $i=1$, the result is obvious from 4.3(iii)
and (13).

Now suppose that $(x_{2,0}^+)^2 x_{1,1}^+.v =0$. By 4.3(ii),
$x_{1,0}^+x_{2,0}^+x_{1,1}^+.v =0$ as well and (13) now forces
$$x_{2,0}^+x_{1,1}^+.v =0.$$ Now, (2)  gives $x_{1,0}^+x_{1,1}^+.v =0$ and so
by  a final application (13), we get
$$x_{1,1}^+.v =0.$$
 One proves similarly that $(x_{2,0}^+)^2x_{1,-1}^+.v =0$ implies that
$x_{1,-1}^+.v =0$, and
the proof  of 4.4 is now complete. \qed\enddemo

We are now in  a position to give the
\demo{Proof of 4.1} All three parts are proved by induction on $r$. If $r=0$,
the result follows from 2.2. We assume that (i), (ii) and (iii)  hold for $r$
and prove them for $r+1$.

\noindent (i) Suppose  that $m_{\lambda-(r+1)\theta}(V(\bp)) >1$. Then, by 4.4,
there exists $0\ne v_0\in V(\bp)_{\lambda-(r+1)\theta}^+$ such that
$x_{1,1}^+.v =0$.

Suppose now that $x_{1,-1}^+.v\ne 0$. For $s=0,1,\ldots ,r+1$, define $v_s\in
V(\bp)$ by
$$v_s =((x_{2,0}^+)^2x_{1,-1}^+)^s.v.$$
We claim that the $v_s$ have the following properties:

\noindent $({\text{i}})_s\ \ 0\ne v_s\in V(\bp)_{\lambda-(r+1-s)\theta}^+$ for
all $0\le s\le r+1$;

\noindent $({\text{ii}})_s\ \  x_{i,k}^+.v_s =0$ for $i=1,2$, $k\ge 0$.

Note that $(\text{i})_0$  holds by assumption and $({\text{ii}})_0$ by the
choice of $v_0$. Assuming that these properties hold for $s$ we now prove that
they hold for $s+1$. Lemma 4.2 implies that $x_{1,-1}^+.v_s\ne 0$ if $0\le s\le
r$ and  4.4 now shows that $v_{s+1}\ne 0$.
To prove that $(\text{ii})_{s+1}$ holds, observe that, by the proof of  4.2, it
suffices to prove that $x_{1,1}^+.v =0$. Using (2) we find that
$$x_{1,1}^+(x_{2,0}^+)^2x_{1,-1}^+\in
 \uqgh x_{1,1}^+x_{2,0}^+x_{1,-1}^+ +\uqgh x_{2,1}^+x_{2,0}^+x_{1,-1}^+ +\uqgh
x_{1,0}^+x_{2,0}^+x_{1,-1}^+.$$
The third term kills $v_s$ by 4.3(ii); on the other hand, using (1), we find
that the first two terms are contained in
$$\sum_{i=1}^2(\uqgh x_{i,0}^++\uqgh x_{i,1}^+),$$
and hence kill $v_s$ as well. This proves the claim.

Note that $v_{r+1} = Av_{\bp}$, for some $A\in\Bbb  C^{\times}$. Since
${\text{dim}}(V(\bp)_{\lambda-\alpha_2}) =1$, it follows that
$$x_{2,0}^+x_{1,-1}^+. v_r = Bx_{2,0}^-.v_{\bp},$$
for some $B\in\Bbb C^{\times}$. Applying $x_{2,1}^+$ to both sides of this
equation, and using (2) and $(\text{ii})_r'$, we get
$$0 = B\phi_{2,1}^+v_\bp.$$
By 1.4(c),  this is impossible, since $\lambda(2)> 0$. This completes the proof
of 4.1(i).

(ii) Suppose that  $m_{\lambda-(r+1)\theta-\alpha_2}(V(\bp)) > 0$.
The induction hypotheses on $r$  implies that
$$ V(\bp)_{\lambda-r\theta-\eta} =0\ \ \text{if} \ \ \eta=\alpha_2, \
2\alpha_2, \ 3\alpha_2, \ \text{or}\ \alpha_2-\alpha_1.\tag14$$
Let $0\ne v\in V(\bp)_{\lambda-(r+1)\theta-\alpha_2}^+$. We shall prove that
$v$ is actually $\uqgh$--highest weight, which is obviously impossible.
We first prove, by induction on $k$, that $x_{1,k}^+.v =0$.
By (14), it suffices to prove that $x_{1,k+1}^+.v\in
V(\bp)_{\lambda-r\theta-3\alpha_2}^+$.
 Since $x_{1,0}^+x_{1,k+1}^+\in\sum_{0\le s<k+1}\uqgh x_{1,s}^+$, we see that
$$x_{1,0}^+.x_{1,k+1}^+.v =0.$$
To prove that $x_{2,0}^+x_{1,k+1}^+.v =0$, define
$v'=(x_{2,0}^+)^2x_{1,k+1}^+.v$ and $v''=x_{1,0}^+.v'$. Then, by (14),
$$\align \text{4.3(iii)}&\implies v''\in
V(\bp)_{\lambda-r\theta-\alpha_2+\alpha_1}^+\implies v'' =0,\\
\text{4.3(i)}&\implies v'\in V(\bp)_{\lambda-r\theta-\alpha_2}^+\implies v'
=0,\\
\text{4.3(ii)}&\implies x_{2,0}^+x_{1,k+1}^+.v\in
V(\bp)_{\lambda-r\theta-3\alpha_2}^+\implies x_{2,0}^+x_{1,k+1}^+.v
=0.\endalign$$

To prove that $x_{2,k}^+.v =0$, we again proceed by induction on $k$. We assume
that $k\ge 0$; the proof for $k\le 0$ is similar.

Using  (1) and the fact that $x_{1,k}^+.v=0$ for all $k$, we see that
$$\align x_{1,r}^+x_{2,k+1}^+.v &=0, \ \ \text{for}\ r=-1,0,\ \text{and}\ 1,\\
x_{2,0}^+x_{2,k+1}^+.v& =0.\endalign$$
But now 4.2 implies that $x_{2,k+1}^+.v =0$ (since
$\lambda-(r+1)\theta\ne\lambda$). This completes the proof of 4.1(ii).

(iii) Let $v\in V(\bp)_{\lambda-(r+1)\theta-\alpha_1-\alpha_2}$. Since
$$m_{\lambda-(r+1)\theta-\alpha_i}(V(\bp))=0, \ \ \text{for}\ \  i=1,2,$$
and since $\lambda\ne\lambda-(r+1)\theta-\alpha_1-\alpha_2$, it suffices by 4.2
 to prove that $x_{2,\pm 1}^+.v =0$. To do this, note that by 3.1, it is enough
 to prove that $x_{2,\pm 1}^+.v\in V(\bp)_{\lambda-(r+1)\theta -\alpha_1}^+$.
Clearly, by (1), $x_{2,0}^+x_{2,\pm 1}^+.v =0$.

To prove that $x_{1,0}^+x_{2,\pm 1}^+.v =0$, it suffices by 4.2 to prove that
$$\align x_{1,0}^+.x_{2,\pm 1}^+.v&\in V(\bp)_{\lambda-(r+1)\theta}^+,\tag15\\
x_{1,s}^+.x_{1,0}^+x_{2,\pm 1}^+. v&=0\ \ \text{for}\ \  s=\pm
1.\tag16\endalign$$
The fact that $(x_{1,0}^+)^2x_{2,\pm 1}^+.v =0$ is clear from (2).
By using (1) and (2), it is easy to see that $x_{2,0}^+x_{1,0}^+x_{2,\pm
1}^+.v\in V(\bp)_{\lambda-r\theta-\alpha_1-\alpha_2}$, and hence must be zero
by 3.1. This proves (15).

To prove (16), one checks first, using (1) and (2), that
$$(x_{2,0}^+)^2x_{1,s}^+x_{1,0}^+x_{2,\pm1}^+\in \uqgh.
x_{2,0}^+x_{1,0}^+x_{2,\pm 1}^++\uqgh.x_{2,0}^+x_{2,\pm 1}^+.$$
It follows that $(x_{2,0}^+)^2x_{1,s}^+x_{1,0}^+x_{2,\pm 1}^+.v=0$ for $s=0,1$.
Lemma 4.4 now implies that  in fact
$$x_{1,s}^+x_{1,0}^+x_{2,\pm 1}^+.v =0\ \text{for}\  s=\pm 1.$$
This completes the proof of 4.1(iii).\ \qed\enddemo

The proof of Theorem 2.4 is now complete.

\vskip 36pt

\noindent {\bf References}
\vskip 12pt

\noindent 1. Beck, J., Braid group action and quantum affine algebras,
preprint, MIT, 1993.

\noindent 2. Chari, V., Minimal affinizations of representations of quantum
groups: the rank 2 case, preprint, 1994.

\noindent 3.  Chari, V. and Pressley, A. N., Quantum affine algebras, Commun.
Math. Phys. {\bf 142} (1991), 261--83.

\noindent 4.  Chari, V. and Pressley, A. N., Small representations of quantum
affine algebras, Lett. Math. Phys. {\bf 30} (1994), 131--45.

\noindent 5.  Chari, V. and Pressley, A. N., {\it A Guide to Quantum Groups},
Cambridge University Press, Cambridge, 1994.

\noindent 6. Chari, V. and Pressley A. N., Minimal affinizations of
representations of quantum groups: the simlpy-laced case, preprint, 1994.

\noindent 7. Chari, V. and Pressley, A. N., Quantum affine algebras and their
representations, preprint, 1994.

\noindent 8. Delius, G.W. and Zhang, Y.-Z., Finite-dimensional representations
of quantum affine algebras, preprint, 1994.

\noindent 9. Drinfel'd, V. G., A new realization of Yangians and quantized
affine algebras, Soviet Math. Dokl. {\bf 36} (1988), 212--6.

\noindent 10. Lusztig, G., {\it Introduction to Quantum Groups}, Progress in
Mathematics 110, Birkh\"auser, Boston, 1993.

\enddocument